# Magnetic properties and spin dynamics in single molecule paramagnets $Cu_6Fe$ and $Cu_6Co$

P.Khuntia[1,2], M.Mariani[1,5], M.C.Mozzati[1], L.Sorace[3], F. Orsini[4,5], A.Lascialfari[1,4,5], F.Borsa[1], M. Andruh[6], C.Maxim[6]

[1]*Dipartimento di Fisica" A.Volta" e Unita' CNISM-CNR, Universita' di Pavia, I-27100 Pavia, Italy*

[2]*Department of Physics, Indian Institute of Technology Bombay, Powai, Mumbai-400076, India*

[3]*Laboratory for Molecular Magnetism and INSTM Research Unit, University of Florence, I-50019 Sesto Fiorentino, Italy*

[4]*Dipartimento di Scienze Molecolari Applicate ai Biosistemi DISMAB, Università di Milano, I-20134 Milano, Italy*

[5] *S3-CNR-INFM, I-41100 Modena, Italy*

[6]*Inorganic Chemistry Laboratory, Faculty of Chemistry, University of Bucharest, Str. Dumbrava Rosie 23, 020464 Bucharest, Romania*

# Abstract

The magnetic properties and the spin dynamics of two molecular magnets have been investigated by magnetization and d.c. susceptibility measurements, Electron Paramagnetic Resonance (EPR) and proton Nuclear Magnetic Resonance (NMR) over a wide range of temperature (1.6-300K) at applied magnetic fields, H=0.5 and 1.5 Tesla. The two molecular magnets consist of $Cu^{II}(saldmen)(H_2O)\}_6\{Fe^{III}(CN)_6\}](ClO_4)_3\cdot 8H_2O$ in short $Cu_6Fe$ and the analog compound with cobalt, $Cu_6Co$. It is found that in $Cu_6Fe$ whose magnetic core is constituted by six $Cu^{2+}$ ions and one $Fe^{3+}$ ion all with s=1/2, a weak ferromagnetic interaction between $Cu^{2+}$ moments through the central $Fe^{3+}$ ion with J = 0.14 K is present, while in $Cu_6Co$ the $Co^{3+}$ ion is diamagnetic and the weak interaction is antiferromagnetic with J = -1.12 K. The NMR spectra show the presence of non equivalent groups of protons with a measurable contact hyperfine interaction consistent with a small admixture of s-wave function with the d-function of the magnetic ion. The NMR relaxation results are explained in terms of a single ion ($Cu^{2+}$, $Fe^{3+}$, $Co^{3+}$) uncorrelated spin dynamics with an almost temperature independent correlation time due to the weak magnetic exchange interaction. We conclude that the two molecular magnets studied here behave as single molecule paramagnets with a very weak intramolecular interaction, almost of the order of the dipolar intermolecular interaction. Thus they represent a new class of molecular magnets which differ from the single molecule magnets investigated up to now, where the intramolecular interaction is much larger than the intermolecular one.

# I) Introduction

The development of molecular chemistry in synthesizing transition metal-ion based molecular clusters whose properties are midway between atoms and bulk systems provides a unique opportunity to the scientific community for the study of nanoscale magnetism [1]. Because of the presence of non-magnetic organic ligands that prevent magnetic interactions, the intermolecular interactions are weak in comparison to intramolecular super-exchange interactions. Hence, the molecules are isolated magnetically from each other and it is of great interest to investigate spin dynamics of these nanomagnets, often called single molecule magnets (SMM). In the SMM reported up to now, a strong exchange magnetic interaction exists among the magnetic moments within each individual molecule which leads to a low temperature ground state characterized by either a high total moment S or a singlet antiferromagnetic (AFM) state S=0 depending on the topology of the magnetic ions and on their mutual magnetic coupling. In the case of high spin ground state and high magnetic anisotropy, quantum tunneling of magnetization and quantum coherence at low temperature have been observed, making these nanomagnets promising candidates for magnetic storage and quantum computing among other applications [2,3].

In this paper we present the magnetic properties of heptanuclear molecular magnets $Cu_6Fe$ and $Cu_6Co$ with very small intramolecular magnetic coupling. These molecules are thus a prototype of single molecule paramagnets. As it will be shown by the experimental results, the $Cu_6Fe$ compound consists of six $Cu^{2+}$ magnetic ions with a weak ferromagnetic interaction via the bond to a central $Fe^{3+}$ ion. The isostructural $Cu_6Co$ compound, instead, appears to be formed by six $Cu^{2+}$ magnetic ions and a central diamagnetic $Co^{3+}$ ion with a small antiferromagnetic coupling between $Cu^{2+}$ ions via the bond to the central $Co^{3+}$ ion.

The paper is organized as follows. In section II we summarize briefly the synthesis of the compounds and their crystal structure. The detailed description of the results of this section will be presented in a separate publication [4]. In Section III we present the experimental results and data analysis. The results are presented in separate subsections for the magnetization, the EPR, the static and the dynamic NMR. Section IV contains a comparison between $Cu_6Fe$ and $Cu_6Co$ and a discussion of the results obtained with the different techniques and the relevant conclusions.

# II) The Samples

## (A) $Cu_6Fe$

Polycrystalline sample$[\{Cu^{II}(saldmen)(H_2O)\}_6\{Fe^{III}(CN)_6\}](ClO_4)_3\cdot 8H_2O$ (i.e, $C_{72}H_{118}C_{13}Cu_6FeN_{18}O_{32}$) was synthesized from the reaction of binuclear copper(II) complex, $[Cu_2(saldmen)_2(\mu\text{-}H_2O)(H_2O)_2](ClO_4)_2\cdot 2H_2O$, with $K_4[Fe(CN)_6]$ (H saldmen is the Schiff base resulted by reacting salicylaldehyde with N,N-dimethylethylenediamine as will be described elsewere [4]). In this molecule, 16 out of 118 protons belong to 8 crystallization water molecules and the remaining 102 protons arise from the organic ligands and from six water molecules co-ordinated to six $Cu^{2+}$ ions.

The lattice is of hexagonal symmetry (*R*-3*c*) with cell constants a=27.8777(16) Å, b=27.8777(16) Å, c=21.369(13) Å, $\alpha=\beta=90^{\circ}$ and $\gamma=120^{\circ}$. The six $Cu^{2+}$ ions are located at the corners of an octahedron and are connected by the cyano groups and one $Fe^{3+}$ at the center of the octahedron. The Cu-Fe-Cu angles are 180° and the Fe-Cu angles across the CN bridges are Cu-N-C=171.76(54)° and Fe-C-N=176.54(57)°.

The nearest neighbor bond distances are Fe-H=4.0482Å and Cu-H=2.9649Å.

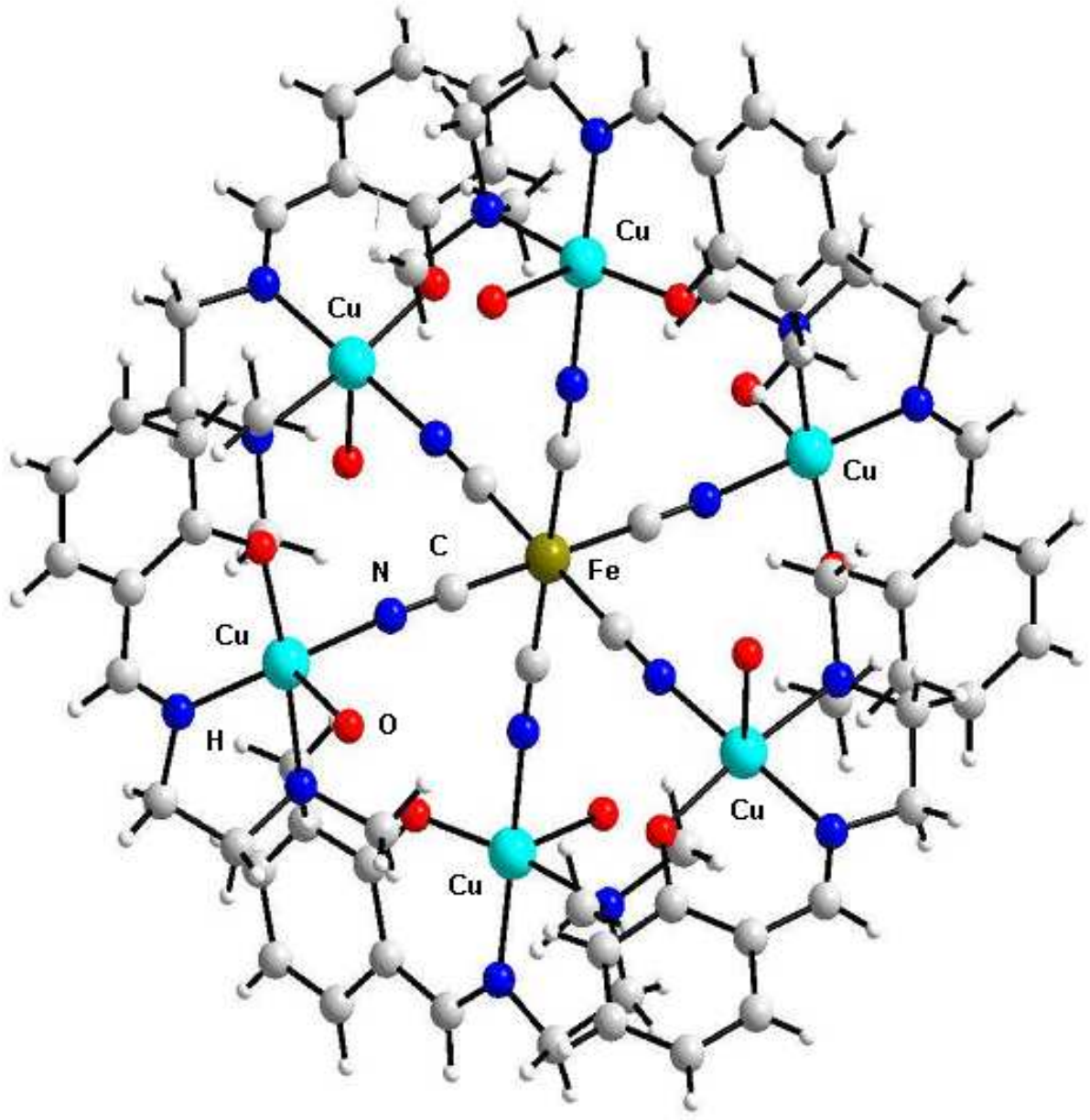

**Fig. 1**. Crystal structure of $Cu_6Fe$ . $Cu_6Co$ is isostructural, with Fe replaced by Co

**(B) <u>$Cu_6Co$</u>**

The $Cu_6Co$ crystals were obtained by adding an acetonitrile-water (1:1) solution (20 mL) containing 0.3 mmol of $[Cu_2(saldmen)_2(\mu\text{-}H_2O)(H_2O)_2](ClO_4)_2\cdot 2H_2O$, 10 mL acetonitrile-water (1:1) solution containing 0.1 mmol of $K_3[Co(CN)_6]$ under stirring. Green crystals suitable for X-ray diffraction were obtained directly from the reaction mixture, by slow evaporation of the filtrate at room temperature [4].

The lattice is also of hexagonal symmetry ($R$-3$c$) with cell constants a=27.9545(19)Å, b=27.9545(19)Å, c=21.3938(16)Å, $\alpha=\beta=90^\circ$ and $\gamma=120^\circ$. The six $Cu^{2+}$ ions are located at the corners of an octahedron and are connected by the cyano groups and one $Co^{3+}$ at the center of the octahedron. The Cu-Co-Cu angles are 180˚ and the Co-Cu angles across the CN bridges are Cu-N-C=174.21˚ and Co-C-N=173.712˚.

The nearest neighbor bond distances are Co-H=3.9836Å and Cu-H=2.9628Å.

# III) Experimental results and analysis

## A. Magnetic susceptibility

The temperature dependence of the magnetic susceptibility ($\chi$=M/H) in the temperature range 2-210 K at two applied magnetic fields, 0.1 Tesla and 1Tesla for $Cu_6Fe$, and in the temperature range 2-160 K at 1Tesla for $Cu_6Co$, was measured with a Superconducting Quantum Interference Device (SQUID) magnetometer. The raw data were corrected by the sample holder and the single ion diamagnetic contributions before analysis.

The results of the susceptibility measurements are shown in fig. 2 for both $Cu_6Fe$ and $Cu_6Co$ samples. Over most of the temperature range the $\chi$T vs T data show a simple paramagnetic behavior. At very low temperature it is evident that a departure from the simple Curie law due to a small ferromagnetic (FM) coupling for $Cu_6Fe$ and a small antiferromagnetic (AFM) coupling for $Cu_6Co$.

The data for $Cu_6Fe$ can be fitted with a Curie-Weiss law with C = 2.72±0.2 (emu.K /mol) and $T_F$ =+0.07 K. This corresponds to an average g factor for the seven spins s=1/2 per molecule of g = 2.035. This is surprisingly low, given the supposedly unquenched orbital momentum characterizing the $^2T_{2g}$ state of low spin Fe(III) in octahedral symmetry which should lead to a much larger average g value [5]. The same behavior has been recently reported for a linear, cyanide bridged, CuFeCu complex, and attributed to the peculiar geometrical distortion of $Fe(CN)_6$ unit, leading to an almost complete quench of the angular momentum [6]. The obtained value of the Weiss constant correspond, in the framework of simple Molecular Field Approximation (MFA), $\theta = \frac{2zs(s+1)J_F}{3k_B}$, to a weak ferromagnetic interaction $J_F$ = 0.14 K.

On the other hand the data for $Cu_6Co$ were fitted with a Curie-Weiss law with C = 2.44±0.06 (emu K /mol) and $T_N$ = -0.56 K. This correspond to six spins s=1/2 with an average g factor g = 2.075. Again, by using the MFA expression for the Weiss constant one finds an antiferromagnetic interaction $J_{AF}$ = -1.12 K.

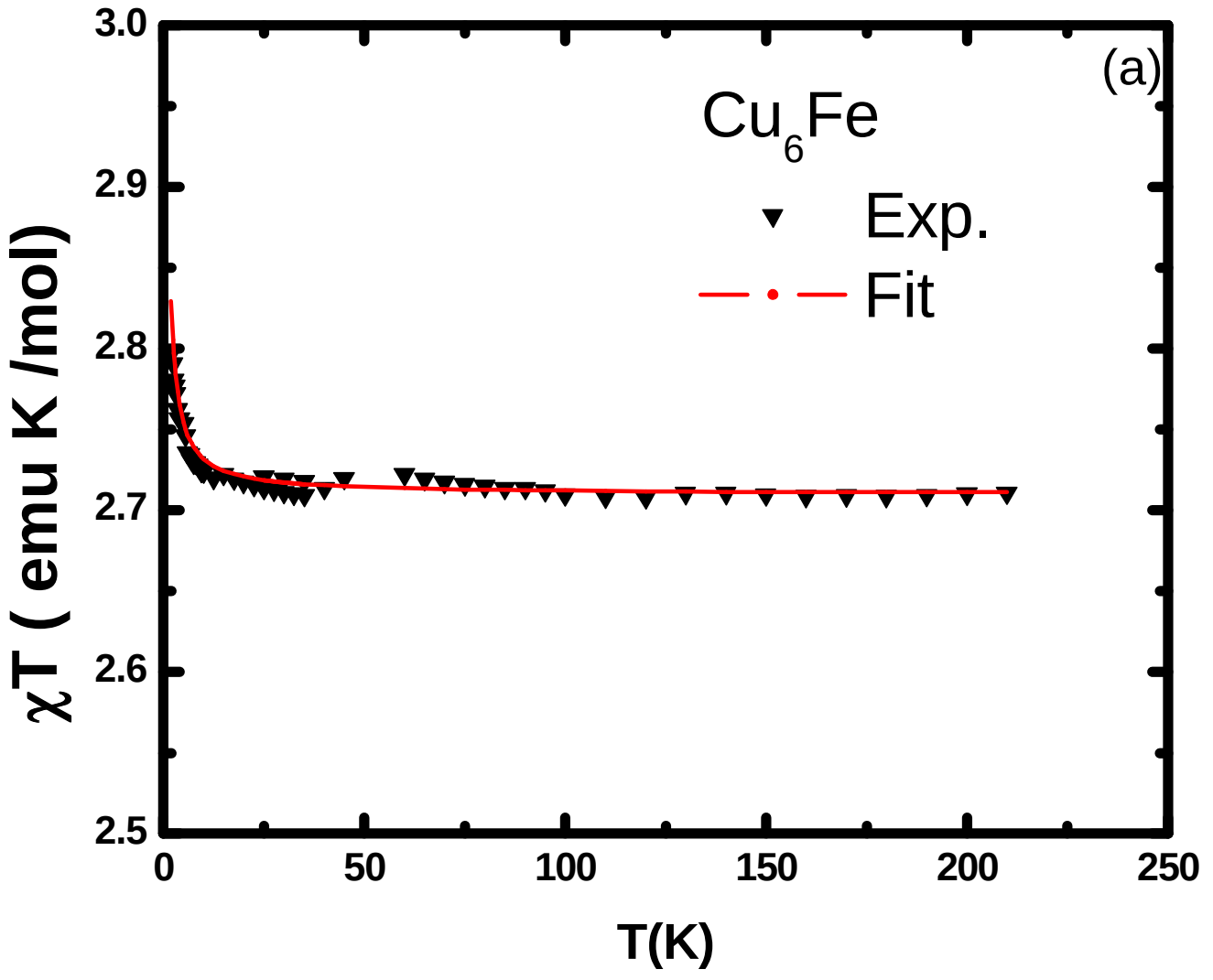

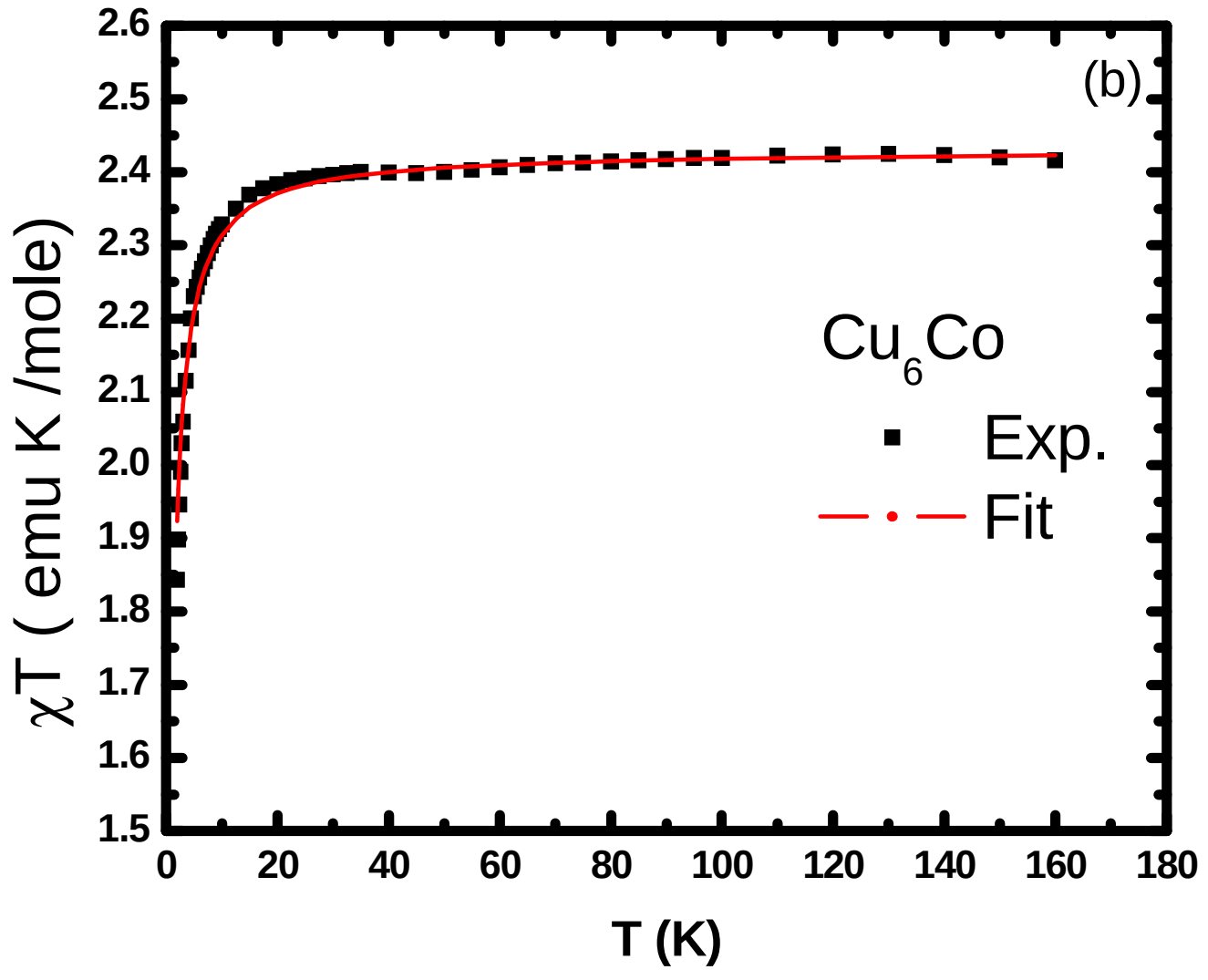

**Fig.2**. Magnetic susceptibility times the temperature vs temperature for (a) $Cu_6Fe$ and (b) $Cu_6Co$. The solid lines are theoretical fits in terms of a Curie-Weiss law as discussed in the text.

As a whole, these results point to the existence of only a very weak exchange coupling interaction between the magnetic centers. This conclusion is reinforced by the isothermal magnetization curves which can be fitted reasonably well in terms of non-interacting paramagnetic ions [4].

While this is not much surprising for the $Cu_6Co$ derivative, for which the interacting centers are located far apart from each other, and mutually counterbalancing interactions may occur, the situation is more puzzling for the $Cu_6Fe$ derivative. For this system the magnetic orbitals of Cu(II) and those of Fe(III), respectively $e_g$ and $t_{2g}$ in octahedral symmetry, should be orthogonal, leading to a substantial ferromagnetic interaction. While the observed interaction is indeed of the expected sign, its magnitude is much lower than expected. It is however to be noted that a negligibly small value of the exchange coupling of Cu(II) with $Fe(CN)_6^{3-}$ units has been recently reported [7].

## B. EPR spectra

Electron Paramagnetic Resonance (EPR) measurements were carried out at 9.45 GHz (X band) at room temperature with a Bruker spectrometer, equipped with a standard microwave cavity. A modulation field of 0.05 mT and a microwave power of about 1.86 mW were used.

The room temperature EPR spectrum of $Cu_6Fe$ and of $Cu_6Co$ are shown in Fig. 3. The shape of the signal for both systems is typical of octahedral $Cu^{2+}$ ions with axial distortion environment, leading to a $g_{//}>g_{\perp}>2.00$ pattern. This is in agreement with the findings of crystal structure solution, which indicated a square pyramidal coordination environment for Cu(II) [8]. The experimental spectrum could be satisfactorily simulated (as a powder spectrum resulting from the superposition of spectra of axial sites with angular orientations randomly distributed) by assuming an anisotropic g-factor with a Lorentzian line shape**.** The values obtained from the simulation of the spectrum are $g_{//}$=2.172 and $g_{\perp}$= 2.085. This confirms that the unpaired electron is located, as expected, in a $d_{x^2-y^2}$ orbital, so that the absence (or weakness) of the exchange coupling between Fe(III) and Cu(II) should be regarded as accidental. Finally, we note that the absence of the EPR signal arising from $Fe^{3+}$ ion in the corresponding derivative at room temperature is most likely due to the fast relaxation time of low spin Fe(III) at this temperature, leading to an exceedingly broad line. Further studies at lower temperatures are currently in progress to clarify this issue.

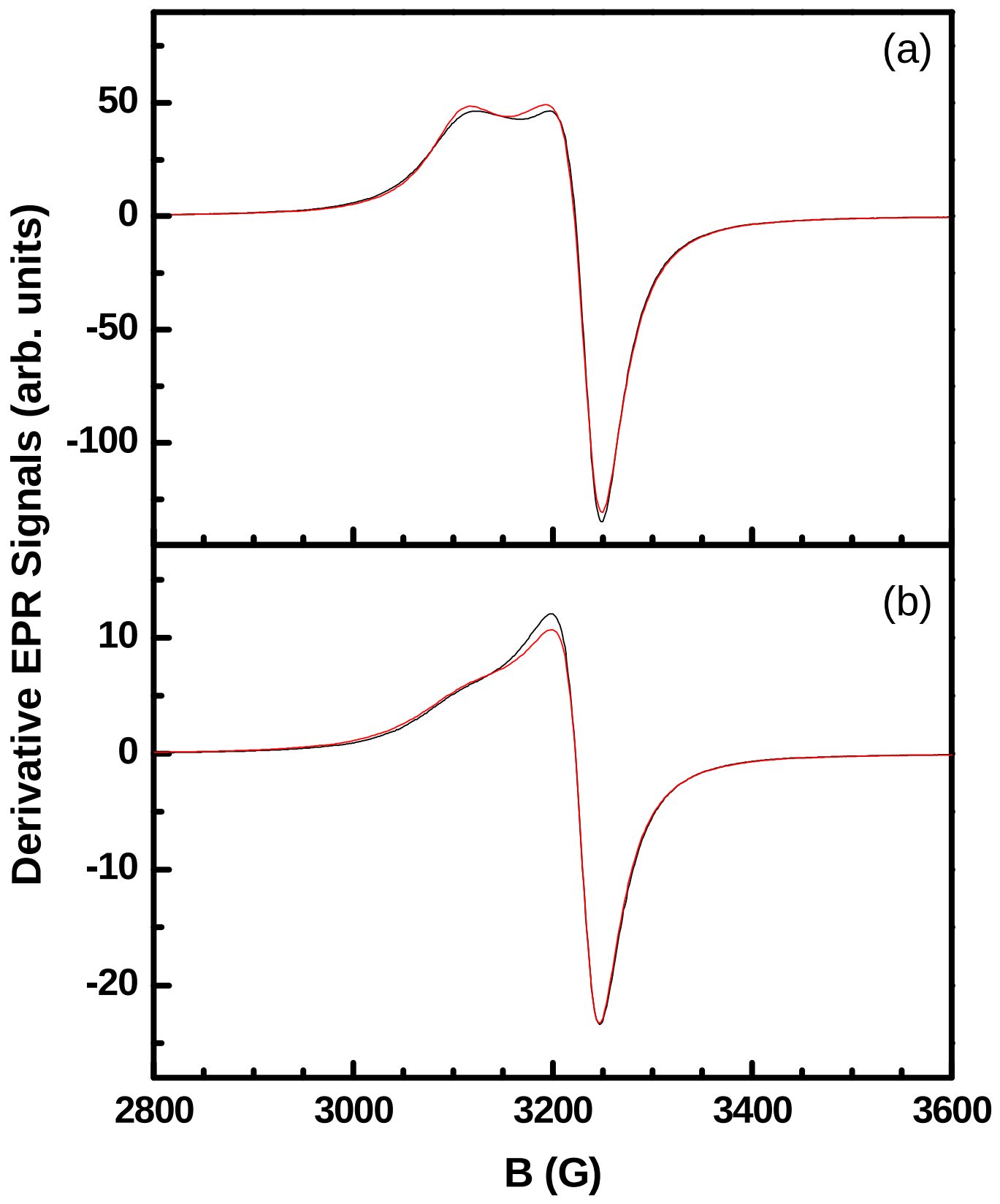

**Fig. 3** Experimental (black line) and computed from numerical analysis (red line) derivative EPR signals in $Cu_6Fe$ (a) and in $Cu_6Co$ (b).

## C. Proton NMR spectra

Nuclear Magnetic Resonance (NMR) measurements on polycrystalline $Cu_6Fe$ and $Cu_6Co$ samples were performed with a standard TecMag Fourier transform pulse NMR spectrometer using short $\pi/2$-$\pi/2$ radio frequency (r.f) pulses (1.9-2.2 μs) in the temperature range 1.6 K to 300 K at two applied magnetic fields, H=0.5 T and 1.5 T. We employed a continuous flow cryostat in the temperature range 4.2 to 300 K and a bath cryostat in the temperature range 1.6 to 4.2 K. Fourier transform of the half echo spin signal of the NMR spectrum was taken in the case where the whole line could be irradiated with one r.f pulse. The low temperature broad spectra were obtained by the convolution of lines obtained from several Fourier transforms each one collected at different values of the irradiation frequency keeping the external field constant.

Proton NMR spectra for $Cu_6Fe$ and $Cu_6Co$ were collected as a function of frequency at constant applied magnetic field H=1.5 T at different temperatures. The spectra thus obtained are shown in Fig.4. They are found to broaden progressively with decreasing temperature and to develop a structure due to the presence of a shifted small component. The spectra at low temperatures could be fitted well with two Gaussian functions having different width and shift. In the analysis of the data which follows we use as experimental results for the full width at half maximum (FWHM) and for the paramagnetic shift $K_{ps} = \frac{\Delta \nu}{\nu_L}$ ($\nu_L$ is the Larmor frequency) the values used for the fitted Gaussian lines.

.

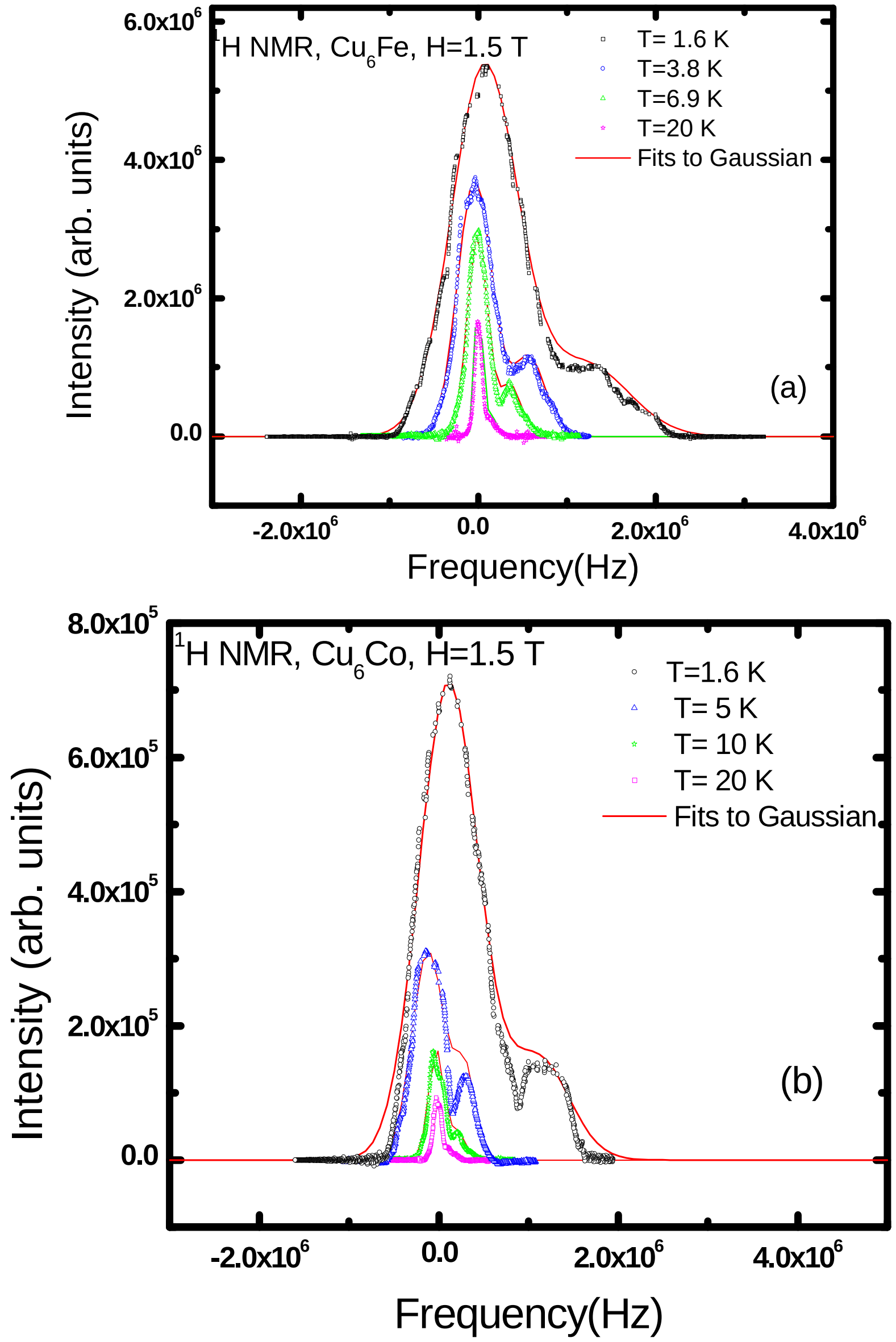

**Fig.4** Representative spectra of proton NMR at different temperatures with fitting curves made up of the superposition of two Gaussian lines at different resonance frequency for $Cu_6Fe$ (**a)** and for $Cu_6Co$ (**b**).

The shape and width of the proton NMR spectrum is determined by two main interactions: (i) the nuclear-nuclear dipolar interaction, (ii) the hyperfine interaction of the proton with the neighboring magnetic ions. The first interaction generates a temperature and field independent broadening [9] which depends on the hydrogen distribution in the molecule and is thus similar in all molecular magnets independently of their magnetic properties [10].

The hyperfine field resulting from the interaction of protons with local magnetic moments of $Cu^{2+}$ may contain contributions from both the classical dipolar interaction and from a direct contact term due to the hybridization of proton s-wave function with the d-wave function of magnetic ions. The dipolar contribution has tensorial character and is thus responsible for the inhomogeneous width of the line due to the random distribution of orientations in a powder sample and to the many non-equivalent proton sites. The contact interaction, on the other hand, has scalar form and it can generate a shift of the line for certain groups of equivalent protons in the molecule [11].

In the usual simple Gaussian approximation for the NMR line shape, the line width is proportional to the square root of the second moment, which in turn is given by the sum of the second moments due to the two interactions described above [9]:

$$FWHM \propto \sqrt{\langle \Delta \nu^2 \rangle_d + \langle \Delta \nu^2 \rangle_m} \qquad (1)$$

where $<\Delta\nu^2>_d$ is the intrinsic second moment due to nuclear dipolar interactions, and $<\Delta\nu^2>_m$ is the second moment of the local frequency-shift distribution (due to nearby electronic moments) at the different proton sites of all molecules. The relation between $<\Delta\nu^2>_m$ and local $Cu^{2+}$ electronic moments for a simple dipolar interaction is given by [9]

$$\langle \Delta \nu^2 \rangle_m = \frac{1}{N} \sum_R \left( \sum_{i \in R} \langle \nu_{R,i} - \nu_0 \rangle_{\Delta t} \right)^2 = \frac{\gamma^2}{N} \sum_R \left[ \sum_{i \in R} \sum_{j \in R} \frac{A(\vartheta_{i,j})}{r_{i,j}^3} \langle m_{z,j} \rangle_{\Delta t} \right]^2 \qquad (2)$$

where R labels different molecules, i and j span different protons and $Cu^{2+}$ ions within each molecule, N is the total number of probed protons. In Eq.2 , $\nu_{R,i}$ is the NMR resonance frequency of nucleus i and $\nu_L = \gamma$ H is the bare Larmor resonance frequency. The difference between the two resonance frequencies represents the shift for nucleus i due to the local field generated by the nearby moments j. A ($\vartheta_{i,j}$) is the angular dependent dipolar coupling constant between nucleus i and moment j and $r_{i,j}$ the corresponding distance. < $m_{z,j}$> is the component of the $Cu^{2+}$ moment j in the direction of the applied field, averaged over the NMR data acquisition timescale. In a simple paramagnet one expects $\langle m_{z,j} \rangle = \frac{\chi}{N_A}$ where $\chi$ is the SQUID susceptibility in emu/mole and $N_A$ is Avogadro's number.

We can thus write approximately:

$$FWHM = \sqrt{\langle \Delta \nu^2 \rangle_m} = A_z \chi \qquad (3)$$

where $A_z$ is the dipolar coupling constant averaged over all protons and all orientations .

The experimental results for the magnetic contribution to the line width are plotted as a function of the magnetic susceptibility in Fig.5 for both $Cu_6Fe$ and $Cu_6Co$. The linear relation predicted by Eq.3 is well verified and the values obtained from the fit for the average dipolar coupling are $A_z = 2.53 \times 10^{22}$ $cm^{-3}$ ( for $Cu_6Fe$ ) and $A_z = 3.44 \times 10^{22}$ $cm^{-3}$ (for $Cu_6Co$) which are consistent with the dipolar interaction of protons not directly coupled to the $Cu^{2+}$ magnetic ions at a mean distance of 3 Å from $Cu^{2+}$.

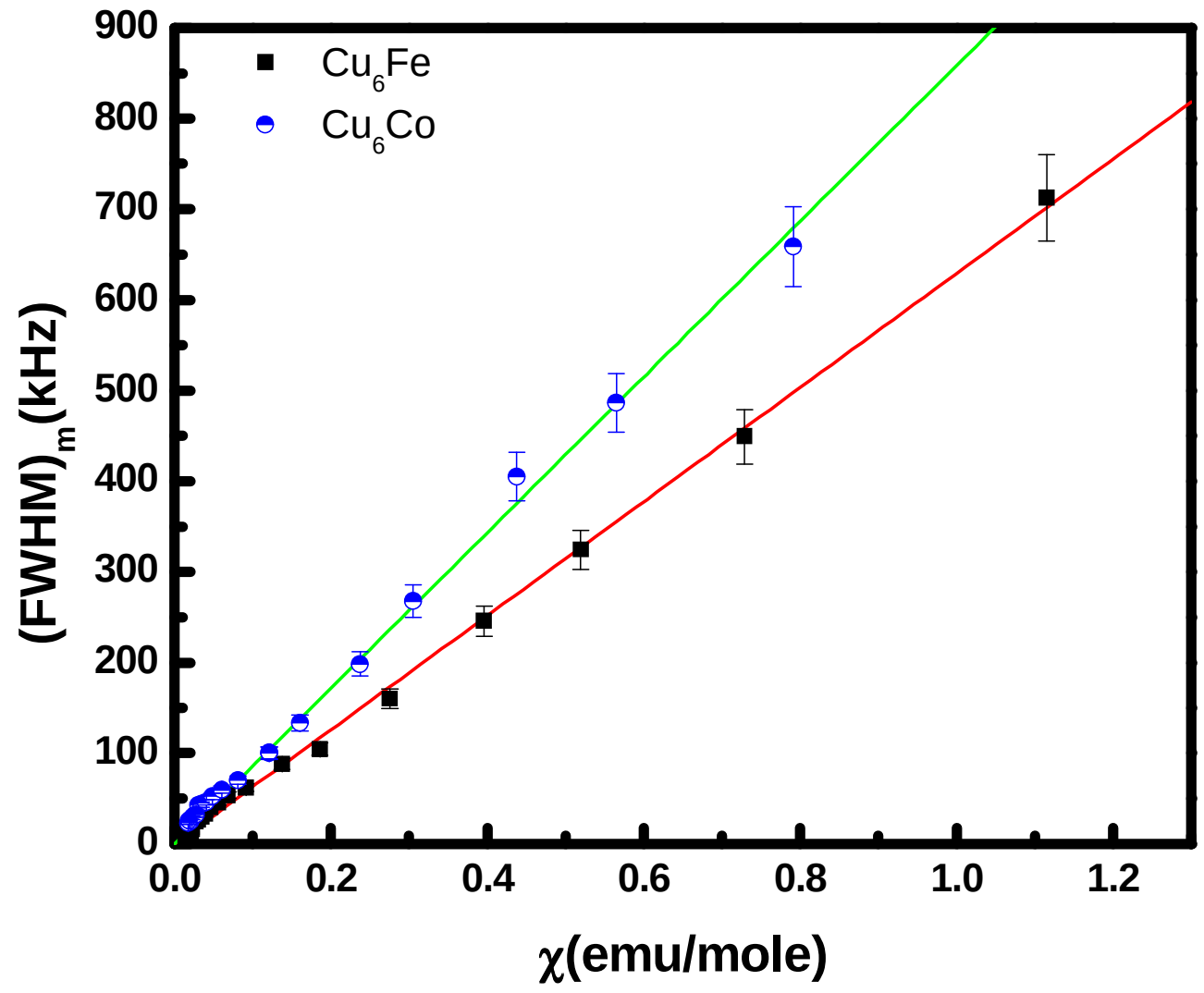

**Fig.5.** Magnetic inhomogeneous broadening of the proton NMR line plotted vs. magnetic susceptibility in $Cu_6Co$ and $Cu_6Fe$. The straight lines are curve fits according to Eq.3 .

We turn now to the analysis of the small shifted line observed in the NMR spectra of both $Cu_6Fe$ and $Cu_6Co$ (see Fig.4). The paramagnetic shift is defined as $K_{ps} = \frac{\nu_R - \nu_L}{\nu_L}$, where $\nu_R$ is the resonance frequency and $\nu_L$ is the proton Larmor frequency . It can be expressed as [11]:

$$K_{ps} = \frac{H_{eff}}{N_A \mu_B} \chi(T) \tag{4}$$

where $\mu_B$= Bohr magneton and $\chi$(T)=paramagnetic susceptibility per mole, $N_A$=Avogadro's number, $H_{eff}$ = local hyperfine field. The hyperfine field, which generates the line shift, is due to a contact scalar interaction arising from the electron density associated with the s- part of the wave function at the proton site. Thus $H_{eff}$ can be expressed in terms of the atomic hyperfine coupling constant, a(s), multiplied by a correction factor, $\xi$, which gives the fraction of s-character of the wave function of the magnetic electron at the proton site [11]:

$$H_{eff} = \mu_B \xi a(s) \qquad (5)$$

For an atom the hyperfine constant can be expressed as $a(s) = \frac{16\pi}{3}\gamma_n \hbar \mu_B P_A$, with $P_A = |\psi_A(0)|^2$ the electron probability density at the nucleus for the free atom.

The experimental results for the shift of the satellite line in Fig.4 are shown in Fig.6 for both $Cu_6Fe$ and $Cu_6Co$ plotted also as a function of the magnetic susceptibility. As seen in the figure the prediction of Eq.4 is well verified. From the slope of the plot of the shift vs. the susceptibility one derives a value of 506.5 G for the hyperfine magnetic field at the proton site for hydrogen bonded to the $Cu^{2+}$ for $Cu_6Fe$ and a value of 462.4 G in the case of $Cu_6Co$.

The theoretical hyperfine constant for H atom is a(s)= 0.0473cm$^{-1}$ [11] close to the value reported for the molecular hydrogen ion $H_2^+$ [12] and corresponding to an hyperfine field at the proton site of about 28 Tesla. Thus the contact term for the bridging hydrogen's in $Cu_6Fe$ and $Cu_6Co$ is only about 0.17% of the atomic hyperfine field for 1s electron in hydrogen atom consistent with a very small overlap of d and s wave functions of the magnetic ion and the hydrogen respectively.

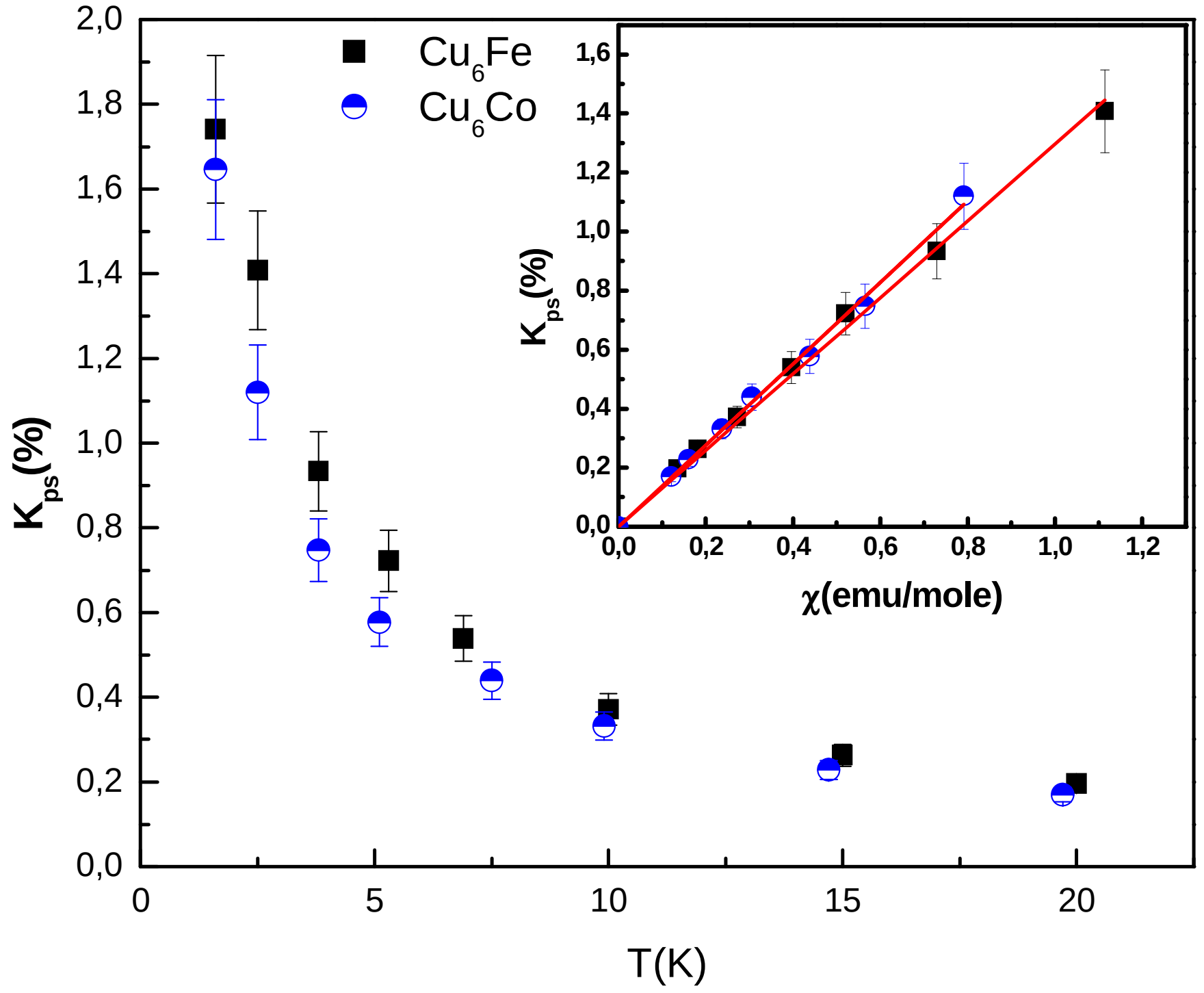

**Fig. 6** Temperature dependence of paramagnetic shift of the satellite line (see Fig.4) in $Cu_6Fe$ and $Cu_6Co$. The inset shows the linear behavior of $K_{ps}$ vs. $\chi$ with temperature as an implicit parameter.

## D. Proton NMR signal intensity, $T_2$ and wipeout effects

The normalised signal intensities for protons studied as a function of temperature in $Cu_6Fe$ and $Cu_6Co$ are shown below. The signal intensity was measured by the area under the echoes collected at different delay times, obtained from then usual Hahn-echo sequence [13]. The $M_{xy}(t)$ vs. t curve giving the spin-spin relaxation recovery law was extrapolated at t=0 and normalised by multiplying by T to compensate for the Boltzmann factor. At low temperature the spectrum broadens and so it was acquired point by point by sweeping the resonance frequency at fixed magnetic field. As shown in Fig.7 the decrease of the normalized intensity in the intermediate temperature regime indicates a loss of signal. The loss of signal is a phenomenon, which has been observed in many molecular nanomagnets [14]. The explanation of this “wipe-out” effect rests in the very short $T_2$ attained by the nuclei closer to the magnetic ions. $T_2$ was measured in our systems from the exponential decay of the echo amplitude as a function of time delay between two rf pulses and the results are shown in Fig. 8. The very short value of $T_2$ and the broad maximum observed in the T dependence of $1/T_2$ in Fig.8 are in qualitative agreement with the loss of signal intensity observed in the same temperature range.

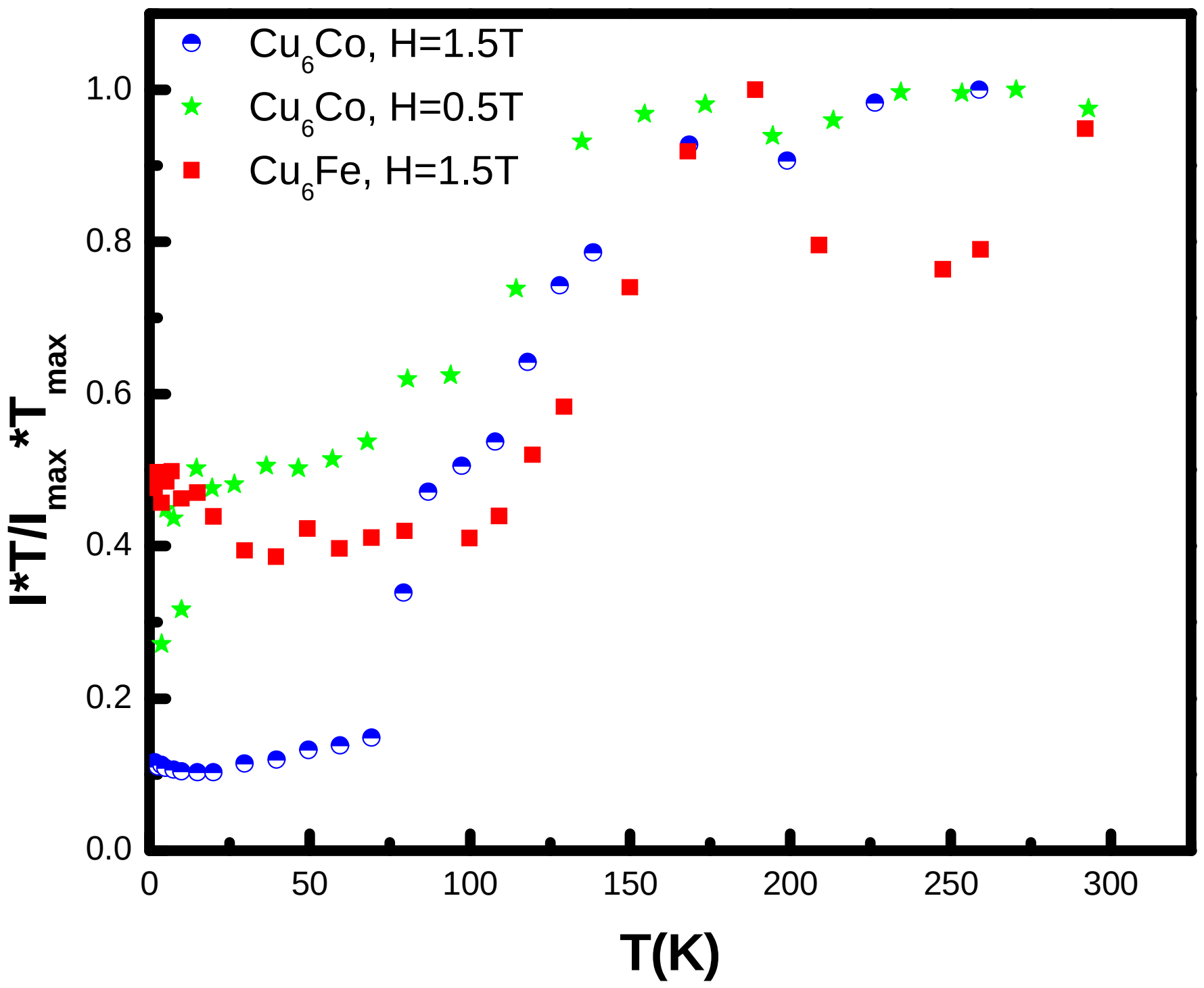

**Fig. 7**. Temperature dependence of normalised NMR signal intensity multiplied by temperature

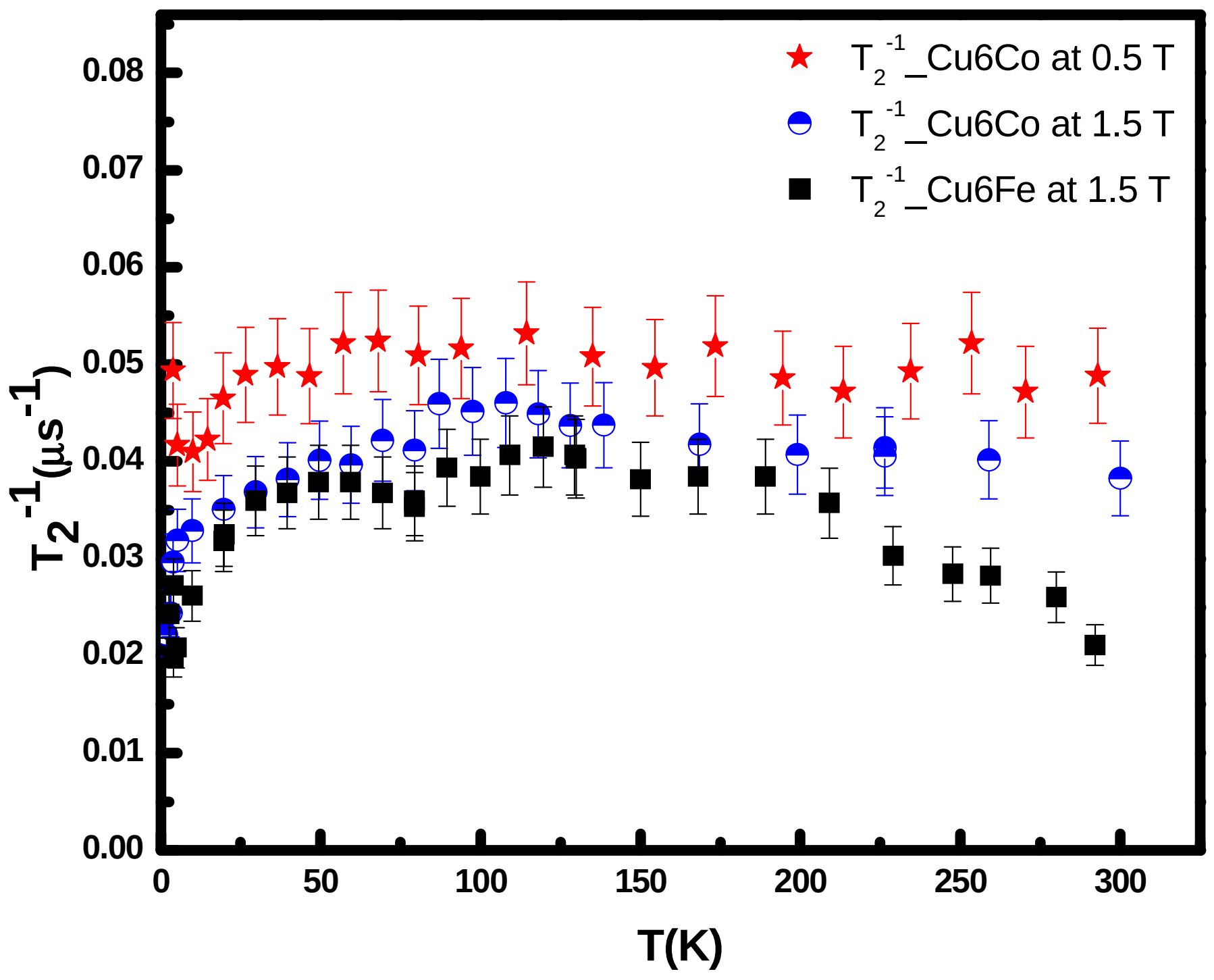

**Fig. 8**. Temperature dependence of spin-spin relaxation rate in $Cu_6Fe$ at H=1.5T and at H=1.5T and at H= 0.5T in $Cu_6Co$.

Thus the results of the temperature and field dependence of the relaxation rate, which will be discussed in the following paragraph, refer only to the average proton relaxation rate of the nuclei which can be detected. Since a large number of nuclei escape detection ( i.e, the above cited "wipeout" effect) the absolute values of $1/T_1$ are clearly not very significant. However, the relative temperature and field dependence should not be affected by the wipe-out effect.

## E. Temperature and field dependence of NSLR

The proton nuclear spin lattice relaxation rate (NSLR), $T_1^{-1}$ , was obtained by monitoring the recovery of the nuclear magnetization following a long comb of $\pi/2$ radio frequency (r.f ) pulses in order to obtain the best initial saturation conditions. The recovery was found to be strongly non exponential at all temperatures and magnetic fields. This is a common situation in molecular nanomagnets [10] since the protons are located at different distances and angles from the relaxing magnetic ions. Since the spin diffusion is not sufficiently fast to ensure a common spin temperature during the recovery process, the recovery curve is a superposition of many exponential curves each one representing the relaxation of a given proton. By measuring the initial recovery or tangent at the origin one measures the average relaxation rate, which is dominated by the fast relaxing protons (the nearest to the magnetic ions). The shape of the recovery curve may change as a function of temperature and magnetic field as the result of the spin diffusion effect [15]. Thus for better consistency we measured the NSLR from the time at which the recovery curve has reduced to 1/e of the initial value. The measured parameter is in any case proportional to the average relaxation rate of the protons detected in the NMR signal [15].

The results for the field dependence of the proton relaxation rate at three different temperatures in both compounds are shown in Fig.9 and 10.

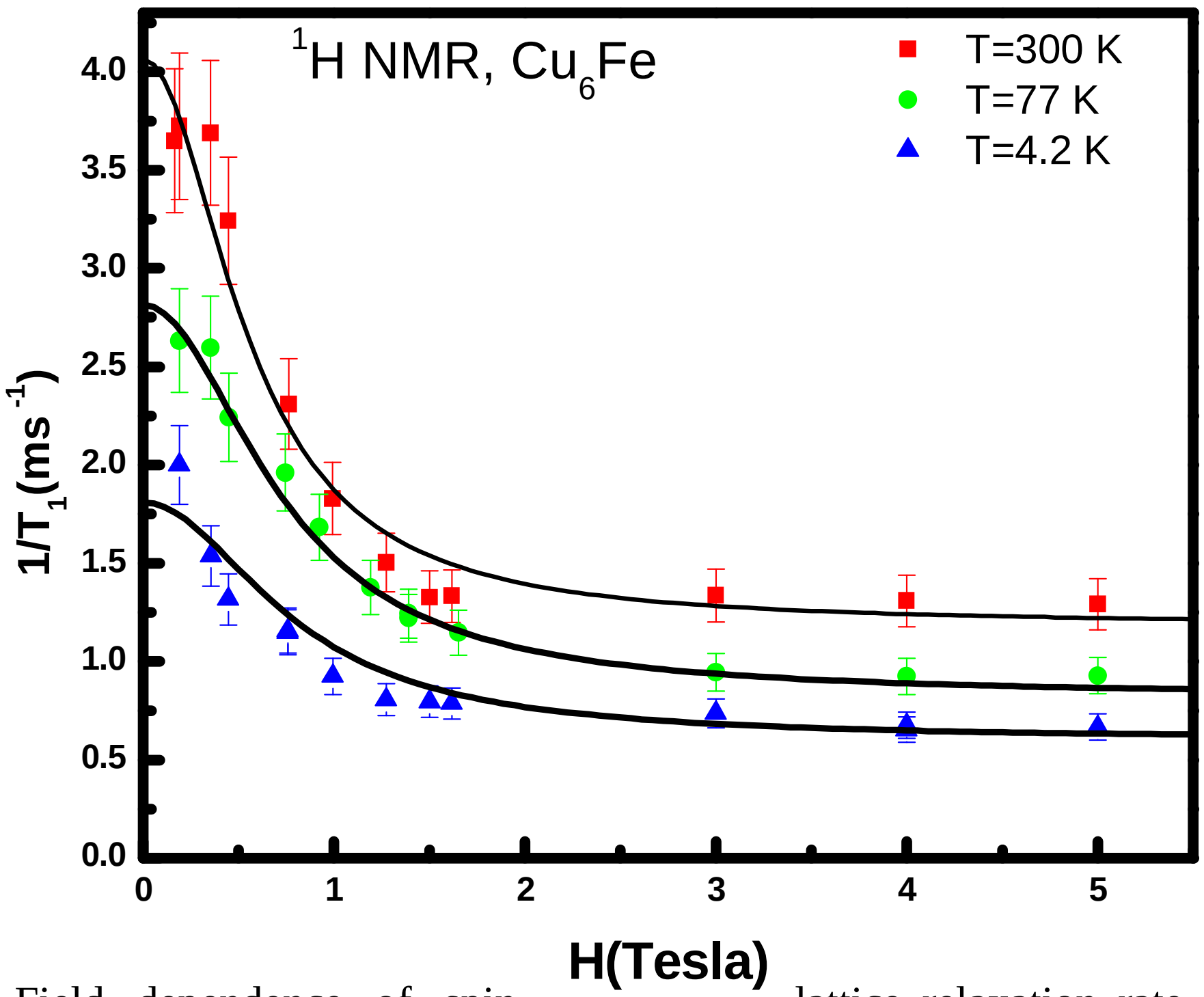

**Fig. 9**. Field dependence of spin-lattice relaxation rate in $Cu_6Fe$ at various temperatures with fit according to Eq(7) ( see text).

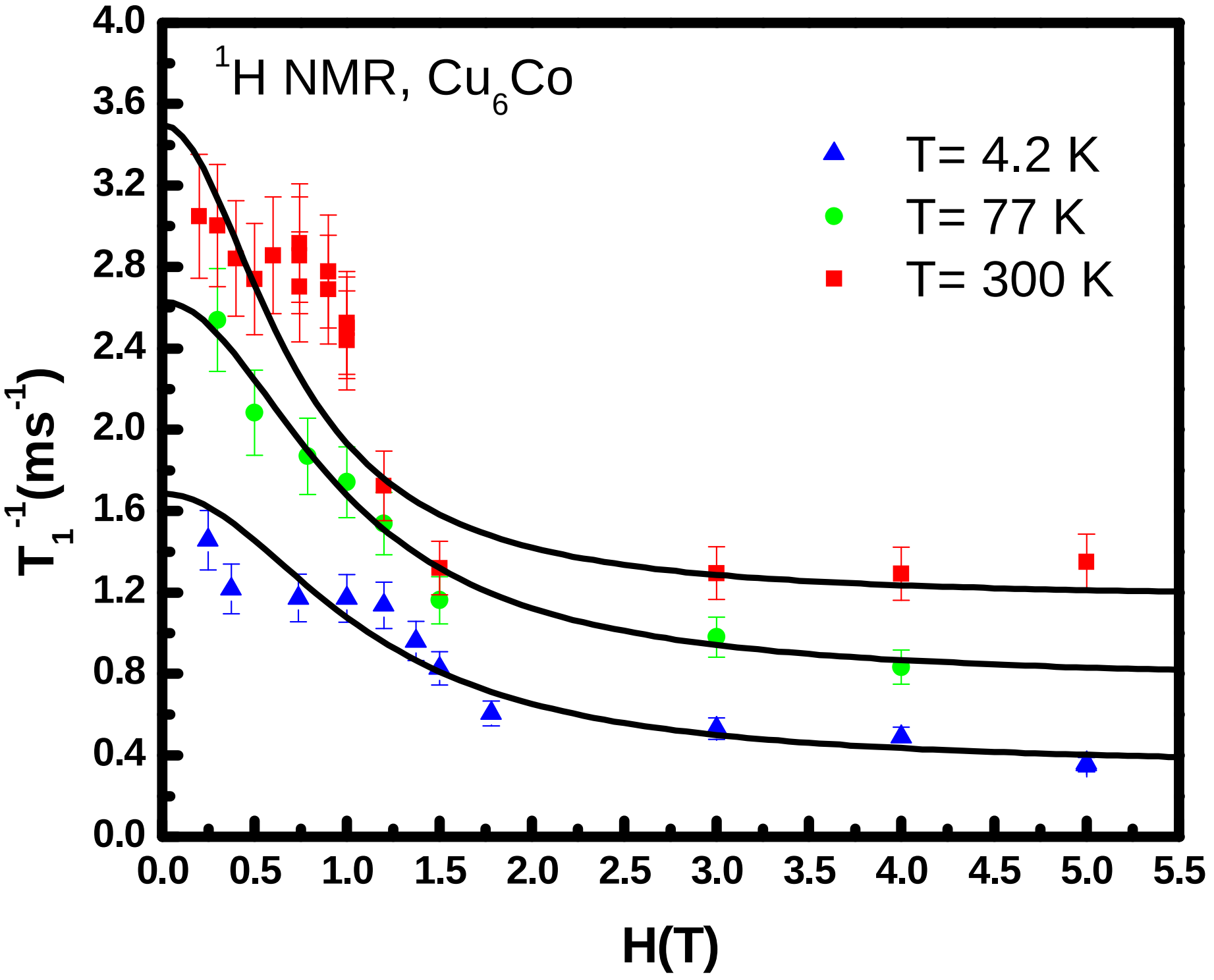

**Fig. 10** Field dependence of spin-lattice relaxation rate in $Cu_6Co$ at various temperatures with fit according to Eq(7). ( see text).

The results for the temperature dependence of proton NSLR at two external magnetic fields, H=1.5 T and H= 0.5 T are shown in Fig.11.

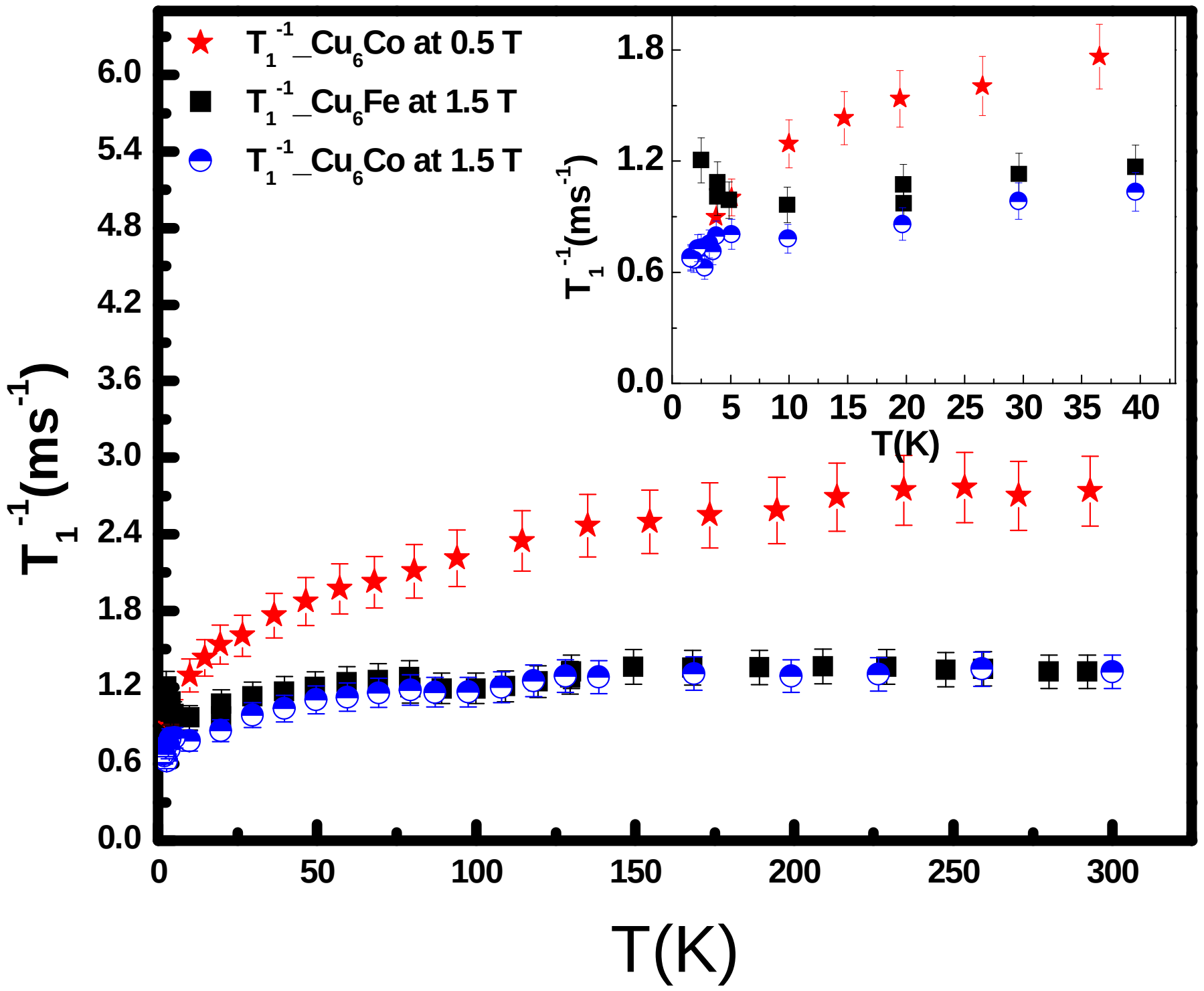

**Fig. 11** Temperature dependence of spin-lattice relaxation rate at H=1.5T in $Cu_6Fe$ and at H=1.5T, 0.5T in Cu6Co. The inset shows the low temperature behavior.

The weak temperature dependence is at variance with the pronounced peak observed in strongly exchange coupled molecular nanomagnets [10]. This is consistent with the simple paramagnetic behaviour observed in the magnetization measurements. One can conclude that the spin dynamics reflects the fluctuations of the single magnetic moments of the ions in the molecule without effects associated to the collective spin dynamics except for the very low T region where an upturn of $1/T_1$ is observed for the FM coupled $Cu_6Fe$ and a downturn of $1/T_1$ is observed in AFM coupled $Cu_6Co$ (see inset of Fig.11). The weak decrease of the relaxation rate in the intermediate temperature range is most likely due to the decrease of spin diffusion time due to the inhomogeneous broadening of the proton NMR line [15].

A more quantitative analysis of the data can be done on the basis of Moriya's theory for NSLR in Heisenberg isotropic three dimensional paramagnets [16,17]

In three dimensional paramagnets the spectral density J (ω) of the electronic spin fluctuation is Lorentzian with a correlation frequency given by [16,17]:

$$\omega_{exc} = \frac{Jk_B}{\hbar}\sqrt{2zs(s+1)} \qquad (6)$$

where z is the number of nearest neighbors and for both the systems $Cu_6Fe$ and $Cu_6Co$, z=1

The NSLR is proportional to the spectral density at both the electronic Larmor frequency $\omega_e$ and at the nuclear Larmor frequency $\omega_N$ [10,16,17]

$$\frac{1}{T_1} = \frac{(\hbar\gamma_e\gamma_n)^2}{4\pi g^2\mu_B^2} k_B T\chi(0)\left[\frac{1}{2}A^{\pm}J^{\pm}(\omega_e) + A^z J^z(\omega_n)\right] \qquad (7)$$

where $A^{\pm}$ and $A^z$ are the Fourier transforms of the spherical component of the product of two dipolar interaction tensors describing the hyperfine coupling of a given proton to the paramagnetic ion along transverse and longitudinal directions, respectively, with respect to the external magnetic field averaged over all protons and all directions [10,16].

In three dimensional paramagnets with strong exchange interaction J the exchange frequency $\omega_{exc}$ is much larger than both $\omega_e$ and $\omega_n$ and thus one finds that the relaxation rate is field independent since $J^{\pm}(\omega_e) = \frac{\omega_{exc}}{\omega_{exc}^2 + \omega_e^2} \cong \frac{1}{\omega_{exc}}$ and $J^z(\omega_n) = \frac{\omega_{exc}}{\omega_{exc}^2 + \omega_n^2} \cong \frac{1}{\omega_{exc}}$. On the other hand in the present case, since the exchange coupling is very small, a field dependence is possible from the first term in Eq.7.

We have fitted the results in Fig.9 and 10 with Eq.7 which can be rewritten for a Lorentzian spectral density and $\omega_{exc} >> \omega_n$ as:

$$\frac{1}{T_1} = K\left[\frac{1}{2}A^{\pm}\frac{\omega_{exc}}{\omega_{exc}^{\ 2}+\omega_e^{\ 2}} + A^z\frac{1}{\omega_{exc}}\right] \qquad (8)$$

The constant K can be estimated from the known value of the susceptibility $\chi(0)$. The values are K =1.7 $ms^{-1}$ for $Cu_6Fe$ and K =1.4 $ms^{-1}$ for $Cu_6Co$. The exchange frequency should be of the order of the value obtained using Moriya's formula [16,17] by using the measured exchange interactions J i.e. $\omega_{exc}$ = $1.3\times10^{10}$ Hz for $Cu_6Fe$ (J= 0.14 K, ferromagnetic) and $\omega_{exc} = 1.036\times10^{11}$ Hz for $Cu_6Co$ ( J= -1.12 K, antiferromagnetic).

The experimental data in Fig.10 and 11 can be fitted by Eq.8 with values for the hyperfine constants: $A^{\pm} \cong 1.4\times10^{46}$ $cm^{-6}$ and $A^z \cong 0.31\times10^{46}$ $cm^{-6}$ for $Cu_6Fe$ and $A^{\pm} \cong 2.07\times10^{46}$ $cm^{-6}$ and $A^z \cong 0.4\times10^{46}$ $cm^{-6}$ for $Cu_6Co$ . The fitting parameters are of the correct order of magnitude as obtained in the case of many other molecular nanomagnets [10,18]. In particular, the values of $A^z$, which depend only on the tensorial dipolar interaction [9], are consistent with a dipolar interaction of protons with nearest neighbor and next nearest neighbor magnetic ions. The coupling constant $A^{\pm}$ , on the other hand, can contain contributions from both the dipolar interaction and the scalar contact hyperfine interaction. In both cases we found that $A^{\pm} > A^z$, which is an indication of the presence of an additional contribution due to a contact interaction arising from the hybridization of hydrogen s wave function with the d wave function of Cu ions as found in the analysis of the spectra ( see section C ). The exchange frequencies which best fits the data are : a) for $Cu_6Fe$ , 1.0 x $10^{11}$ Hz , 1.3 x $10^{11}$ Hz and 1.4 x $10^{11}$ Hz at 300K, 77K and 4.2K respectively b) for $Cu_6Co$ , 1.2 x $10^{11}$ Hz , 1.7 x $10^{11}$ Hz, 1.9 x $10^{11}$ Hz for 300K,77K and 4.2 K respectively. In both cases the estimated error is ± 10% . The weak temperature dependence is probably irrelevant since the recovery of the nuclear magnetization and thus the value of the measured NSLR can be affected in a different way at different temperatures by spin diffusion effects which are too difficult to account for. The order of magnitude of the exchange frequency extracted from the data is in excellent agreement with the theoretical value from Moriya's Eq.6 ($\omega_{exc} = 1.036\times10^{11}$ Hz ) only for $Cu_6Co$ . For $Cu_6Fe$ the experimental value is one order of magnitude larger.

This could be due to a much faster fluctuation rate for the $Fe^{+3}$ magnetic moment as also suggested by the impossibility of detecting the EPR signal.

## IV) Summary and Conclusions

We have shown that $Cu_6Fe$ and $Cu_6Co$ are novel magnetic molecular clusters in the sense that, contrary to most of the molecular nanomagnets [1], the magnetic centres are very weakly coupled within the cluster. Thus a crystal of $Cu_6Fe$ (Co) is made up of identical single molecule paramagnets. In $Cu_6Fe$ the $Cu^{2+}$ ions (s=1/2) are found to be coupled in pairs via the magnetic $Fe^{3+}$ (s=1/2) ion by a super-exchange ferromagnetic interaction with $J_F$ = 0.14 K. In view of the weakness of the coupling constant it could also be a simple dipolar coupling between the $Cu^{2+}$ ion and the $Fe^{3+}$ ion. On the other hand in $Cu_6Co$, the $Cu^{2+}$ ions appear to be coupled via the diamagnetic $Co^{3+}$ ion by a super-exchange antiferromagnetic interaction with $J_{AF}$ =-1.12 K.

In both compounds the EPR spectra are indicative of an octahedral $Cu^{2+}$ site with axial distortion (i.e. $g_{//}$=2.172 and $g_{\perp}$= 2.085). The proton spin–lattice relaxation time is consistent with an almost temperature independent single correlation frequency $\omega_{exc}$ = $10^{11}$ Hz related to the fluctuations of the electron spin due to the $T_2$ –type flip-flop transitions associated to the weak exchange coupling as predicted by Moriya [16,17]. However, a quantitative disagreement with the simple Moriya's prediction is found for $Cu_6Fe$ for which we cannot find the EPR signal of the $Fe^{3+}$ ion a circumstance which suggest a fast fluctuation of this ionic magnetic moment not accounted for by Moriya's theory which applies to isotropic Heisenberg interactions only. Below about 2 K the proton relaxation rate shows an increase in $Cu_6Fe$ due to the ferromagnetic correlations and a decrease in $Cu_6Co$ due to antiferromagnetic correlations (see inset Fig.11). Measurements at much lower temperature are necessary to investigate the possible presence of long range magnetic order.

In conclusion the magnetic molecular clusters studied here appear to be very suitable candidates to investigate magnetic ordering at very low temperature ( millikelvin range) where the competition between the weak intramolecular exchange interaction and the even weaker intermolecular dipolar interaction may lead to some novel kind of magnetic order.

## Acknowledgements

We acknowledge support from the EU Network of Excellence MAGMANet and from Grant PRIN N.2006029518 of the Italian Ministry of Research.

---